\input harvmac
\input epsf
\def\np#1#2#3{Nucl. Phys. B {#1} (#2) #3}
\def\pl#1#2#3{Phys. Lett. B {#1} (#2) #3}
\def\plb#1#2#3{Phys. Lett. B {#1} (#2) #3}

\def\physrev#1#2#3{Phys. Rev. D {#1} (#2) #3}

\def\ev#1{\langle#1\rangle}

\def\O{{\cal O}} 
\def\Ot{{\cal O}_{\tau}^{(-4)}}
\def\Otbar{{\cal O}_{\overline \tau}^{(4)}} 
\def\f#1#2{\textstyle{#1\over #2}}

\def\K{{\cal K}}
\def\d{\delta}
\def\cd{d}

\nref\malda{J.M. Maldacena, hep-th/9711200, Adv. Theor. Math. Phys. 2
(1998) 231.}
\nref\GKP{S.S. Gubser, I. Klebanov, and A.M. Polyakov, hep-th/9802109,
\pl{428}{1998}{105}.}
\nref\EW{E. Witten, hep-th/9802150, Adv. Theor. Math. Phys. 2 
(1998) 253.}
\def\adsr{\refs{\malda - \EW}}
\nref\FMMR{D.Z. Freedman, S.D. Mathur, A. Matusis, and L. Rastelli,
hep-th/9804058, \np{546}{1999}{96}.}
\nref\LMRS{S. Lee, S. Minwalla, M. Rangamani, and N. Seiberg,
hep-th/9806074, Adv. Theor, Math. Phys. 2 (1998) 697.} 
\nref\DFS{E. D'Hoker, D.Z. Freedman, and W. Skiba, 
hep-th/9807098, \physrev{59}{1999}{045008}.}
\nref\GRPS{F. Gonzalez-Rey, I. Park, K. Schalm, hep-th/9811155,
\pl{448}{1999}{37}.}
\nref\EHSSW{E. Eden, P.S. Howe, C. Schubert, E. Sokatchev, P.C. West, 
hep-th/9811172.}
\nref\DFMMR{E. D'Hoker, D.Z. Freedman, S.D. Mathur, A. Matusis,
L. Rastelli, hep-th/9903196.}
\nref\BGKR{M. Bianchi, M.B. Green, S. Kovacs, G. Rossi,
hep-th/9807033, JHEP 9808 (1998) 013.}
\nref\DHKMV{N. Dorey, T.J. Hollowood, V.V. Khoze, M.P. Mattis,
S. Vandoren, hep-th/9901128.}
\def\instr{\refs{\BGKR,\DHKMV}}
\lref\FPZ{S. Ferrara, M. Porrati, A. Zaffaroni, hep-th/9810063.}
\lref\AFGJ{D. Anselmi, D.Z. Freedman, M.T. Grisaru, A.A. Johansen,
hep-th/9708042, \np{B526}{1998}{543}.}
\lref\FPZ{S. Ferrara, M. Porrati, A. Zaffaroni, hep-th/9810063.}
\lref\GMZZ{M. Gunaydin, D. Minic, and M. Zagermann, hep-th/9810226.}
\lref\KI{K. Intriligator, hep-th/9811047, Nucl. Phys. B, to appear.}
\lref\AFGJ{D. Anselmi, D.Z. Freedman, M.T.Grisaru, and A.A.
Johansen, hep-th/9608125, \plb{394}{1997}{329}; hep-th/9708042,
\np{526}{1998}{543}.}  
\lref\ferr{L. Andrianopoli and S. Ferrara, hep-th/9803171; 
\pl{430}{1988}{248}; S. Ferrara, M.A. Lledo, and A. Zaffaroni,
hep-th/9805082, \physrev{58}{1998}{105029}.}
\lref\GM{M. Gunaydin and N. Marcus, Class. Quant. Grav. 2, (1985) L11.}
\lref\BJEF{M. Baker and K. Johnson, Physica A 96 (1979) 120;
J. Erlich and D. Freedman, hep-th/9611133, \physrev{55}{1997}{6522}.}
\lref\BGKRi{T. Banks and M. Green, hep-th/9804170, JHEP 9805 
(1998) 002; 
R. Kallosh and A. Rajaraman, hep-th/9805041,
\physrev{58}{1998}{125003}.}
\lref\harmrefs{P Howe and P. West, hep-th/9509140; hep-th/9607060,
\pl{389}{1996}{273}; hep-th/9611074; hep-th/9611075, 
\pl{400}{1997}{307}, hep-th/9808162, \pl{444}{1998}{341}.}
\lref\Park{J.H. Park, hep-th/9903230.}

\Title{hep-th/9905020,  UCSD/PTH-99/06, IASSNS-HEP-99/45}
{\vbox{\centerline{Bonus Symmetry and the Operator Product Expansion} 
\centerline{of ${\cal N}=4$ Super-Yang-Mills
}}}
\medskip
\centerline{Kenneth Intriligator and Witold Skiba}
\vglue .5cm
\centerline{UCSD Physics Department}
\centerline{9500 Gilman Drive}
\centerline{La Jolla, CA 92093}

\bigskip
\noindent

The superconformal group of ${\cal N}=4$ super-Yang-Mills has two
types of operator representations: short and long.  We conjecture that
operator product expansions for which at least two of the three
operators are short exactly respect a bonus $U(1)_Y$ R-symmetry, which
acts as an automorphism of the superconformal group.  This conjecture
is for arbitrary gauge group $G$ and gauge coupling $g_{YM}$.  A
consequence is that $n\leq 4$-point functions involving only short
operators exactly respect the $U(1)_Y$ symmetry, as has been
previously conjectured based on AdS duality.  This, in turn, would
imply that all $n\leq 3$-point functions involving only short
operators are not renormalized, as has also been previously
conjectured and subjected to perturbative checks.  It is argued that
instantons are compatible with our conjecture.  Some perturbative
checks of the conjecture are presented and $SL(2,Z)$ modular
transformation properties are discussed.

\Date{5/99}           
\newsec{Introduction}

The central objects which characterize conformal field theories in any
dimension are the spectrum of operator dimensions and the operator
product expansion (OPE) coefficients.  These objects control the
behavior of operator correlation functions, which are the observables
of conformal field theories.  Over the past several years, it has been
appreciated that four dimensional gauge theories with enough matter
generically lead to interacting conformal field theories, and it is an
interesting avenue in field theory to consider correlation functions
in any of these theories.

This avenue has not been much explored until recently, in the context
of the maximally supersymmetric 4d conformal field theory, ${\cal
N}=4$ super-Yang-Mills.  The motivation behind the recent work is the
conjectured duality of \adsr\ to gravity in anti-de Sitter space.
There have been recent studies, {}from both the ${\cal N}=4$ field
theory and AdS gravity dual perspectives, of various $n$-point
functions; see, e.g. \refs{\GKP - \DHKMV} and references cited
therein.

We will here conjecture that the OPE coefficients of ${\cal N}=4$
supersymmetric Yang-Mills {\it exactly} obey certain selection rules.
As will be discussed, a motivation for this conjecture comes {}from the
conjectured $AdS$ duality.  Nevertheless, our conjecture itself is
purely a statement about the ${\cal N}=4$ field theory, and thus
logically separate {}from the $AdS$ duality conjecture; it should be
possible to prove or disprove it purely in the context of field
theory.  Indeed, we believe that our conjecture is correct for ${\cal
N}=4$ super-Yang-Mills with arbitrary gauge group $G$.  Unfortunately,
we have not been able to prove the conjecture, so we will here only be
able to present its motivation and some checks within the context of
instantons and perturbative ${\cal N}=4$ field theory.

As reviewed in the next section, operators form representations of the
${\cal N}=4$ superconformal group, which has two types of
representations: the generic ``long'' representations, and the special
``short'' representations.  The short representations are the
generalizations of chiral superfields and analogs of BPS states and
satisfy special properties thanks to supersymmetry; for example, their
dimensions are not renormalized.  It is the short representations
which are seen as single particle states in the $AdS$ supergravity
dual.  All operators in a short representation will be referred to as
``short operators,'' while those in a long representation will be
referred to as ``long operators.''

As emphasized in \KI, the ${\cal N}=4$ superconformal algebra admits a
bonus $U(1)_Y$ symmetry, which acts on the supersymmetry generators as
an $R$-symmetry.  (See also \refs{\GM, \FPZ} for earlier discussions
of $U(1)_Y$.) Although $U(1)_Y$ is not a symmetry of the field theory,
all operators can be assigned a definite $U(1)_Y$ charge and, based on
the $AdS$ duality, it was conjectured in \KI\ that all correlation
functions of short operators, for $G=SU(N)$ (and also $SO(N)$,
$Sp(N)$, and theories with less SUSY obtained by orbifolds) {\it
approximately} respect a $U(1)_Y$ conservation selection rule in the
double limit where $g_{YM}^2N\gg 1$ and $g _{YM}^{-2}N\gg 1$, where
the supergravity dual is weakly coupled.

It was further conjectured in \KI\ that the $U(1)_Y$ selection rule is
actually {\it exact} for correlation functions of $n\leq 4$ short
operators\foot{Generally, there can be contact term contributions to
correlation functions, involving delta functions which vanish unless
operators are at the same point, which violate these selection rules.
For example, in ${\cal N}=4$ supersymmetric $U(1)$ gauge theory we
have $\ev{F_{\alpha \beta}(x)F^{\alpha \beta}(y)}\sim \delta ^4
(x-y)$, where the contact term violates the selection rule by $4$
units.  We will always take the operator insertion points to be
separated and thus ignore contact terms.}.  We believe that this
statement applies for ${\cal N}=4$ with arbitrary gauge group $G$ and
gauge coupling $g_{YM}$.  On the other hand, it is known that the
$U(1)_Y$ selection rule is definitely violated for general $n\geq 5$
point functions of short operators.

The above conjecture, that a selection rule is exact for $n\leq 4$
point functions of operators, but generally violated for $n\geq 5$
point functions, prompts the question: ``why should $n\leq 4$ point
functions be so different {}from $n\geq 5$ point functions?''  In fact,
we point out that there is indeed a natural difference between $n\leq
4$ and $n\geq 5$ within the context of the OPE; this is the motivation
for our conjecture.  The statement of our conjecture is that the OPE
coefficients involving either three short operators $(SSS)$ or two
short and one long operator $(SSL)$ exactly respect the $U(1)_Y$
symmetry.  On the other hand, OPE coefficients involving more than one
long operator ($SLL$ or $LLL$) generally violate the $U(1)_Y$
symmetry.  As we discuss, this conjecture has as a consequence that
the $U(1)_Y$ symmetry is exact for $n\leq 4$ point functions but
violated for $n\geq 5$ point functions of short operators.  The exact
$U(1)_Y$ selection rule for $n\leq 4$ point functions, in turn,
implies \KI\ the non-renormalization of 3-point functions of short
operators which was conjectured in \LMRS\ and checked in the weakly
coupled field theory limit to leading order in perturbation theory in
\DFS.

An outline of this paper is as follows.  In the next section, we
review the representation theory of the ${\cal N}=4$ superconformal
group and its $U(1)_Y$ outer automorphism.  In sect.\ 3, we
discuss the OPE and the motivation for our conjecture.  In sect.\ 4 we
discuss how the supercharges act on gauge-invariant, composite,
operators and demonstrate in perturbation theory that it can be
necessary to include quantum corrections to descendent operators in
the case of long representations of the superconformal group.

In sect.\ 5, we discuss instanton contributions to operator
correlation functions.  We present a heuristic argument for a simple
relation between the $U(1)_Y$ charge of an operator and how many of
the 16 exact fermion zero modes which it contains.  Since instantons
contribute to correlation functions only if the operators saturate all
16 of the exact fermion zero modes, the relation we present leads to
selection rules for when instantons can contribute to correlation
functions.  For an $n$-point function of all short operators,
instantons or anti-instantons can contribute only if $|q_T|=4(n-4)$,
where $q_T$ is the net $U(1)_Y$ charge violation.  This is nicely
consistent with the conjectured non-renormalization of $n\leq 3$ point
functions \refs{\LMRS, \DFS}, and the conjecture \KI\ that the
$U(1)_Y$ selection rule is exact for $n=4$-point functions.  For
correlation functions of two short and one long operator, we argue
that instantons can contribute only if $q_T=0$, which is compatible
with our conjecture that three-point functions with two short and one
long operator exactly vanish if $q_T\neq 0$.  Finally, for correlation
functions of two long operators we argue that instantons only
contribute if $q_T=0$, which is compatible with our argument that all
two-point functions of either short or long operators exactly respect
the $U(1)_Y$ selection rule.

Sect.\ 6 presents a non-trivial, perturbative field theory check of
the conjecture \KI\ that $n\leq 4$ point functions of short operators
respect the $U(1)_Y$; leading order radiative corrections which could
have violated $U(1)_Y$ ``miraculously'' sum to zero, much as in the
3-point functions analyzed in \DFS.  This result applies for any gauge
group $G$. 
  
In sect.\ 7 we discuss perturbative field theory checks of our OPE
conjecture involving two short and one long operator.  In a variety of
examples, we verify that leading contributions to correlation
functions which would violate our conjectured selection rule indeed
sum to zero.  In many cases, this vanishing is a simple consequence of
a sum over color indices $a,b$ of the form $f^{abc}d_{(ab);c_i}$,
which vanishes because $f^{abc}$ is antisymmetric in $[ab]$ while
$d_{(ab);c_i}$ is symmetric.  In many other cases, the required
vanishing of leading order radiative corrections is much more
difficult to verify and it was beyond our patience to complete the
task.  We also present an example where the quantum correction found
in sect.\ 4 for a long descendent is precisely correct to ensure that
a possible violation to our selection rule is indeed canceled.  On the
whole, we find our checks presented in sect.\ 7 to be somewhat
disappointing in that we did not find many tractable examples with the
sorts of ``miraculous'' cancellations found in \DFS\ and our sect.\ 6.
On the other hand, at least all tractable examples are indeed
consistent with our conjectured OPE selection rule, even if the
vanishing is not so impressive.

Finally, in sect. 8, we make some comments regarding $SL(2,Z)$
$S$-duality, the OPE, and our conjectures.

Of course, as with the conjectured non-renormalization of 3-point
functions of short operators \refs{\LMRS, \DFS}, instantons and
leading order perturbative checks can provide some confidence that
conjectured exact statements are indeed correct, but they are no
substitute for a proof.  It is quite possible that our conjecture
concerning the OPE can actually be proven by making full use of the
powerful constraints of the ${\cal N}=4$ superconformal Ward
identities.  Indeed, because all superconformal primary operators are
neutral under $U(1)_Y$, the non-trivial content of our conjecture is a
statement regarding correlation functions involving superconformal
descendents.  If, as naively expected (but it is not at all obvious if
it's true), all superconformal descendent correlation can be obtained
via Ward identities {}from correlation functions involving only
superconformal primary operators, it {\it must} be possible to prove
or disprove our conjecture directly via superconformal Ward
identities.  We have not yet found such a direct proof and leave this
as a problem for future investigation\foot{The ${\cal N}=4$ harmonic
superspace formalism of \harmrefs\ is designed to efficiently make use
of the superconformal Ward identities.  However, we are presently wary
of this formalism as it is purely on-shell and was shown in \KI\ to
lead to the {\it incorrect} conclusion that {\it all } $n$-point
functions of short operators exactly respect the $U(1)_Y$ selection
rule, for all $n$.  (See also \Park.)  As pointed out in \KI, it is
possible that the formalism can be salvaged by finding some missing
superconformal invariants which violate $U(1)_Y$, though this remains
to be seen.  In any case, it does not seem well suited for including
operators in long representations.}.

As mentioned above, we believe that our conjecture applies exactly for
${\cal N}=4$ with any gauge group $G$.  This is in line with the
expectation that it is actually a consequence of supersymmetric
Ward-identities.  If true, this implies that the exact $U(1)_Y$ 
selection rule for $n\leq 4$ point functions and the
non-renormalization of $n\leq 3$ point functions of short operators
also apply for arbitrary gauge group $G$ and gauge coupling $g_{YM}$.  
It is indeed possible to verify that the cancellations of radiative
corrections found in \DFS\ in the context of $G=SU(N)$ occur for
arbitrary gauge group $G$:  the only group theory identity needed in
\DFS\ was $[[T^a,T^p],T^p]=NT^a$, which is a statement about the
quadratic Casimir and generalizes to arbitrary group $G$ as 
$[[T^a,T^p],T^p]=C_2(G)T^a$, with $C_2(G)$ the quadratic Casimir of
the adjoint representation.

\newsec{${\cal N}=4$ superconformal reps. and 
the $U(1)_Y$ bonus symmetry}

The 4d ${\cal N}=4$ superconformal group $PSU(2,2|4)$ has two types of
representations, which we refer to as ``short'' and ``long''.  All
representations are generated by a primary operator $\O _P$, along
with descendent operators, related to $\O _P$ by supersymmetry, of the
form $\delta ^n\overline \delta ^m \O _P$, and their conformal
descendents.  Here $\delta ^n\overline \delta ^m \O _P$ denotes a
nested graded commutators with the sixteen supercharges $Q^I_{\alpha}$
and $\overline Q_{I, \dot \alpha}$, e.g. $\d ^2\overline \d\O _P\equiv
[Q, \{Q,[\overline Q,\O _P]\}]$.  (The remaining 16 superconformal
supercharges act as lowering operators.)  For the generic, long,
representation, the operators $\delta ^n\overline
\delta ^m \O _P$ truncate at $n\leq 8$ and $m\leq 8$ by Fermi
statistics.  All of the operators $\delta ^n\overline
\delta ^m \O _P$ in the long multiplet are referred to as ``long
operators.''

The short representations have the defining property that they instead
truncate at $n\leq 4$ and $m\leq 4$; they are the analog of BPS
objects of superconformal field theories.  It turns out that such
representations are completely characterized by an integer $p\geq 0$.
In particular, the dimensions of such operators are fixed in terms of
$p$ and thus not renormalized.  The spectrum of short representations
was found in \GM\ and their table can also be found reproduced in 
\KI.  The primary operators which generates the short
representations are $\O _p \sim [\Tr _G(\phi ^p)]_{(0,p,0)}$, where
$\phi$ is the ${\cal N}=4$ scalar in the adjoint of the gauge group
$G$ and the ${\bf 6}=(0,1,0)$ of the $SU(4)_R$ global symmetry and
$(0,p,0)$ are the Dynkin labels of the $SU(4)_R$ representation.
There are rank$(G)$ independent short representations (in addition to
the identity), labeled by $p$ which are the degrees of the independent
Casimirs of $G$.  We refer to all operators $\delta ^n\overline \delta
^m \O _p$ in short representation multiplets as ``short operators.''

A simple example of a long representation primary operator is the
$SU(4)_R$ singlet $\O _\K \sim [\Tr _G (\phi ^2)]_{(0,0,0)}$.  As
discussed in \ferr, the multiplet of long operators associated with
$\O _K$ includes the ``Konishi current,'' which is discussed
extensively in \AFGJ. More generally, we can obtain long, primary
operators as $[\Tr _G (\phi ^{r+2s})]_{(0,r,0)}$ via taking a
completely symmetric combination of the $\phi$s and then taking any
number $s>0$ traces.  Operators which are not completely symmetric in
the $\phi$s are descendents since, in $N=1$ SUSY notation, $\overline
D^2
\overline \Phi ^i=g\epsilon ^{ijk}[\Phi _j,\Phi _k]$, where
($\overline \Phi ^i$) $\Phi _i$ are $N=1$ (anti) chiral superfields
and $i=1,2,3$. Other examples of long representations are multi-trace
operators, with more than a single trace over the gauge indices.

In \KI\ it was emphasized that $PSU(2,2|4)$, admits an outer
automorphism $U(1)_Y$, which acts as an $R$ symmetry, under which the
supercharges transform with charge $\pm 1$.  The $U(1)_Y$ charge
assignment of the short representations was determined in the original
analysis of \GM\ in the context of the 5d, ${\cal N}=8$, $AdS_5$
supergravity which is dual
\adsr\ to the 4d ${\cal N}=4$ field theory.  In terms of the field
theory, all operators can be assigned definite charges under $U(1)_Y$;
we write each operator as $O_i^{(q_i)}$, indicating the $U(1)_Y$
charge $q_i$.  The charge assignment is determined as follows
\refs{\GM, \KI}: the adjoint scalar $\phi$ of ${\cal N}=4$ is assigned
charge zero, while the supercharges $Q_\alpha ^I$ and $\overline
Q_{\dot \alpha,I}$ are assigned charges $-1$ and $+1$, respectively.
Thus all primary representations $O_P(x)$, whether for a short or a
long multiplet, which have the form of a trace or traces of
symmetrized powers of $\phi$, are assigned charge zero.  The
superconformal descendents of the form $\delta ^n\overline \delta ^m O
_P$ have $U(1)_Y$ charge $m-n$.  All operators in short
representations thus have $U(1)_Y$ charges $|q|\leq 4$, while all
operators in long representations have charges $|q|\leq 8$.

$U(1)_Y$ is generally {\it not} a symmetry of the ${\cal N}=4$ field
theory.  Nevertheless, it was argued in \KI\ (see also \FPZ) that
$U(1)_Y$ is an {\it approximate} ``bonus symmetry'' of general
correlation functions of small representation operators in an
appropriate limit ($g_{YM}^2N\gg 1$ with also $g_{YM}^{-2}N\gg 1$).
It was further conjectured to be an {\it exact} (i.e. valid for all
gauge groups and all $g_{YM}$) symmetry of all $n\leq 4$ point
functions of operators in small representations.  The argument of \KI\
relied on the AdS duality: $U(1)_Y$ is an approximate symmetry of
$IIB$ string theory in its classical supergravity limit.  However, by
arguments similar to those of \BGKRi, it was suggested in \KI\ that
the stringy and quantum corrections to supergravity, which generally
violate $U(1)_Y$, in fact vanish for all $n\leq 4$ point functions of
short operators.  While $U(1)_Y$ is conjectured to be an exact
symmetry of $n\leq 4$ point functions, it definitely can {\it not} be
an exact symmetry of general $n\geq 5$ point functions of operators in
small representations.  This is seen directly \refs{\BGKR - \KI} both
in the field theory, as will be reviewed below, and via the AdS
duality.

\newsec{Bonus symmetry and the OPE}

Our interest is in characterizing the extent to which the operator
product expansions respect the $U(1)_Y$ bonus symmetry.  We will show
that the OPE is indeed compatible with $U(1)_Y$ being an exact
symmetry of $n\leq 4$ point functions of operators in short
representations, but generally violated for $n\geq 5$ point functions.

We will be interested in operator product expansions of the general
form (for simplicity we write expressions for scalar operators)
\eqn\ope{O_i(x)O_j(y)\sim \sum _k C_{ij}^k
{1\over (x-y)^{\Delta _i+\Delta
_j -\Delta _k}}O_k (y),} with some OPE coefficients $C_{ij}^k$ which
can in general depend on the choice of gauge group $G$, the gauge
coupling $g_{YM}$, and $\theta _{YM}$.  The OPE coefficients appearing
in \ope\ also appear in the three-point functions
\eqn\tpfn{\ev{O _i(x)O _j(y)O_k(z)}={C_{ijk}\over (x-y)^{\Delta
_i+\Delta _j -\Delta _k}(y-z)^{\Delta _j+\Delta _k-\Delta
_i}(x-z)^{\Delta _i+\Delta _k-\Delta _j}}.}  Indices are lowered as
$C_{ijk}=\sum _mC_{ij}^m\eta _{km}$, with the metric $\eta _{ij}$ given
by
\eqn\metric{\ev{O _i(x)O _j(y)}={\eta _{ij}\over (x-y)^{2\Delta _i}},}
which, by conformal invariance, satisfies $\eta _{ij}=0$ unless
$\Delta _i=\Delta _j$.

Any of the operators appearing in \ope, \tpfn, and \metric\ can be
either of short or long type.  Unless indicated otherwise, the indices
$i$, $j$, and $k$ above run over all operators of both types.  When we
want to restrict attention to an operator of a given type, we use
superscripts to denote the type, e.g. $C^{(SSS)}_{ijk}$ when $i$, $j$,
and $k$ are each taken to run only over short operators,
$C^{(SSL)}_{ijk}$ when $i$ and $j$ are short but $k$ is long, etc. 

Consider first the metric \metric.  The correlation function is
non-zero only if the operators $O_i$ and $O_j$ have the same anomalous
dimension, conjugate $SO(4)$ Lorentz spins, and conjugate $SU(4)_R$
representations.  These requirements prohibit a non-zero two-point
function with one operator in a short representation and the other in
a long representation, 
\eqn\metsl{\eta ^{(SL)}_{ij}=0.}  The reason is that
operators are in a short representation if and only if their
dimensions are correlated with their $SU(4)_R$ transformation
properties; if $O_i$ is such a short representation and $O_j$ must
have the same dimension and conjugate $SU(4)_R$ representation as
$O_i$, then $O_j$ must also be in a short representation.  It was
proven in \KI\ that the two-point function of operators in short
representations exactly obey the $U(1)_Y$ selection rule.
The argument of \KI\ did not rely on the operators being
in short representations and we expect that the selection rule
applies for long representations as well:
\eqn\metricYc{\eta ^{SS}_{ij} =0 \quad \hbox{and}\quad \eta
^{LL}_{ij}=0 \quad \hbox{unless}\quad q_i+q_j=0.}  In short, 
the metric $\eta _{ij}$ exactly respects the $U(1)_Y$
selection rule for all operators.

We now consider three point functions and $U(1)_Y$ bonus symmetry.
There are generally non-zero three-point functions involving any
combinations of short and long operators: $C^{(SSS)}_{ijk}$,
$C^{(LLL)}_{ijk}$, $C^{(SSL)}_{ijk}$, and $C^{(LLS)}_{ijk}$. Our main
conjecture is that all three-point functions involving at least two
short representations exactly respect the $U(1)_Y$ selection rule:
\eqn\SSSYc{C^{(SSS)}_{ijk}=0 \quad \hbox{unless} \quad q_i+q_j+q_k=0;}
\eqn\SSLYc{C^{(SSL)}_{ijk}=0 \quad \hbox{unless} \quad q_i+q_j+q_k=0.}
On the other hand, as discussed below, we know that there are
\eqn\LLSYc{\hbox{some}\quad C^{(LLS)}_{ijk}\neq 0\quad  
\hbox{with}\quad q_i+q_j+q_k\neq 0;}
i.e. the case $(LLS)$ of two longs and a short generally does {\it
not} respect the $U(1)_Y$ selection rule.  Similarly, the case $(LLL)$
of three longs is generally {\it not} expected to obey the $U(1)_Y$
selection rule.

As emphasized in \KI\ there is a special short representation
operator: the exactly marginal operator $\Ot$, corresponding to
changing the gauge coupling $\tau ={\theta _{YM} \over 2\pi}+4\pi
ig_{YM}^{-2}$; as indicated, this operator carries $U(1)_Y$ charge
$q=-4$.  There is a conjugate operator $\Otbar$, corresponding to
changing $\overline \tau$.  The metric $\eta _{\tau \overline
\tau}\neq 0$ (it's proportional to $|G|$) and $\eta _{\tau \tau}=\eta
_{\overline \tau \overline \tau}=0$ thanks to \metricYc.  The
variation of a general $n$-point correlation function with respect to
the gauge coupling $\tau$ is given in terms of the $n+1$-point
function with an insertion of $\int \Ot$:
\eqn\corrd{\partial _\tau \ev{\prod _{i=1}^n
O ^{(q_i)}_i(x_i)}=\tau _2^{-1}
\int d^4z\ev{
\Ot (z)\prod _{i=1}^nO ^{(q_i)}_i(x_i)}.}
Note that the correlation function on the left side has total $U(1)_Y$
charge $\sum _{i=1}^nq_i$, while that on the right side has total
$U(1)_Y$ charge $-4+\sum _{i=1}^nq_i$.  If both sides were required to
respect the $U(1)_Y$ selection rule, both sides would have to vanish:
the $n$-point function would be independent of $\tau$ for all $\tau$
-- i.e. be not renormalized.  

A consequence of the $U(1)_Y$ selection rule for two-point and
three-point functions involving all short operators is thus that
two-point functions of operators in short representations are not
renormalized.  It's known that the dimensions of short representations
can not be renormalized, so the content of this statement is that the
metric $\eta ^{(SS)}_{ij}$ in \metric\ is also not renormalized,
i.e. it is independent of $g_{YM}$ and $\theta _{YM}$.  This agrees
with the vanishing of the leading order, radiative corrections found
in \DFS.

The reason why we {\it know} that the $U(1)_Y$ selection rule {\it
must} be violated for general $C_{ijk}^{(LLS)}$ \LLSYc\ is that we
know that operators in long representations generally {\it do} receive
quantum corrections to their anomalous dimensions.  For example, as
shown in
\AFGJ, the ``Konishi current'' $J^\mu _\K$, which is a descendent of
the long primary operator $\O _K$ mentioned above, gets a non-zero
radiative correction to its anomalous dimension.  The operator $\O _K$
and the current $J^\mu _K$ both carry $U(1)_Y$ charge zero.  Since
$\partial _\tau
\ev{J _\K ^\mu (x)J _\K ^\nu (y)}\neq 0$, \corrd\ gives $\ev{\Ot
(z)J_\K ^\mu (x)J _\K ^\nu (y)}\neq 0$, 
which is a $C^{(SLL)}_{ijk}$ which
violates the $U(1)_Y$ selection rule.

As shown in \AFGJ, the operator product expansion of two stress
tensors, which are short operators, includes the Konishi current,
which is a long operator.  Since the stress tensor and the Konishi
current both have vanishing $U(1)_Y$ charge, this is compatible with
our conjecture that all non-zero $C^{(SSL)}_{ijk}$ exactly respect the
$U(1)_Y$ selection rule.

Consider now four-point functions of operators in short
representations.  We assume that there is an expansion of the four
point function in terms of the OPEs of the form
\eqn\ssss{\ev{\prod _{i=1}^4O_i(x_i)^{(q_i)}}=\sum
_{j}C^{(SSX)\ j}_{12}C^{(XSS)}_{j34}F_{\{ i\} ;j}(x_i),} where $X$
denotes that $j$ should be summed over all representations, both short
and long, and we will not be concerned with the form of the functions
$F_{\{ i\} ;j}(x_i)$.  A consequence of
\SSSYc, \SSLYc, and \metricYc\ is that the right side of \ssss\
exactly vanishes unless the charges of the operators satisfy the
$U(1)_Y$ selection rule.  Our OPE conjectures thus implies the
conjecture of \KI\ that, for a general 4-point function of operators
in short representations,
\eqn\SSSSYc{\ev{\prod _{i=1}^4 \O_i ^{(q_i)}(x_i)}=0 \quad
\hbox{unless}\quad \sum _{i=1}^4 q_i=0.}  A consequence of this exact
selection rule for 4-point functions is that all three-point functions
of operators in short representations are not renormalized, as
explained above; i.e. the $C^{(SSS)}_{ijk}$ are constants, independent
of $g_{YM}$ and $\theta _{YM}$.  

We now turn to five-point functions of operators in short
representations.  Again, assuming that an OPE expansion is valid,
these will be of the form
\eqn\sv{\ev{\prod _{i=1}^5 O_i ^{(q_i)}(x_i)}=\sum
_{j,k}C^{(SSX)\ j}_{12} C^{(XSY)\ k}_{j3}C^{(YSS)}_{k45}F_{\{ i\}
;j;k}(x_i).}  Unlike the above case of four-point functions, the OPE
for two longs and a short representation enters as $C^{(LSL)\ k}_{j3}$
in the expansion \sv.  Because these OPE violate the $U(1)_Y$
selection rule, the 5-point function \sv\ of short representations
does {\it not} satisfy an exact selection rule, i.e. it is generally
possible to have
\eqn\svno{\ev{\prod _{i=1}^5O _i^{(q_i)}(x_i)}\neq 0
\quad\hbox{with}\quad \sum _{i=1}^5 q_i\neq 0.}
This situation clearly generalizes for higher $n\geq 5$ point
functions.  

The violations \svno\ of $U(1)_Y$ for $n\geq 5$ point functions can be
seen in the context of the field theory, for example in instanton
contributions to correlation functions \instr .  Instantons will be
discussed further in sect. 5.  There are also contributions to \svno\
which violate the $U(1)_Y$ for $n\geq 5$ point functions which are
visible in perturbation theory.  For example, the perturbative
renormalization of 4-point functions demonstrated in
\refs{\GRPS,\EHSSW} implies via
\corrd\ a perturbative violation of the $U(1)_Y$ selection rule for
the 5-point function with an additional insertion of $\Ot$.
Violations of $U(1)_Y$ for $n\geq 5$ point functions is also
compatible with $AdS$ duality, where it is associated with the stringy
corrections to supergravity.

\newsec{The form of descendent operators}

Because superconformal primary operators are neutral under $U(1)_Y$,
the non-trivial content of the $U(1)_Y$ selection rule conjectures is
for correlation functions involving at least one superconformal
descendent operator.  It is thus important to determine the correct
form of the descendent operators $\delta ^n \overline \delta ^m O_P$.
The on-shell supersymmetry transformations of the ${\cal N}=4$
Yang-Mills theory are given by
\eqn\susyvar{\eqalign{\cd A_{\alpha \dot \alpha}&= \overline \eta
^{I\dot \beta}\epsilon _{\dot \alpha \dot \beta} \psi _{I\alpha}+\eta
^\beta _I\epsilon _{\alpha \beta}\overline \psi ^I_{\dot \alpha}\cr
\cd \phi _{[IJ]}&=\eta ^\alpha _{[I}\psi _{J]\alpha}+\epsilon
_{IJKL} \overline \eta ^{K\dot \alpha}\overline \psi ^L_{\dot
\alpha}\cr 
\cd \psi _{I\alpha}&=\eta ^\beta _IF_{(\alpha \beta )}+\overline
\eta ^{J\dot \beta}\partial _{\alpha \dot \beta}\phi _{IJ}
+g \eta_J^\beta \epsilon_{\alpha \beta} [\phi_{IK},\phi^{JK}] \cr
\cd F_{(\alpha \beta )}&=\overline \eta ^{I\dot \gamma }\partial
_{\dot \gamma (\alpha }\psi _{\beta )I},}} where $\eta ^\alpha _I$ and
$\overline \eta ^{I\dot \alpha}$ are Grassmann parameters to keep
track of the action of $Q_{\alpha}^I$ and $\overline Q _{I \dot
\alpha}$, there are similar transformations for 
$\overline \psi ^l_{\dot
\alpha}$ and $\overline F_{\dot \alpha \dot \beta}$, 
and we have left out numerical constants for simplicity.    
The variations under the other 16
superconformal supersymmetries $S_{\alpha I}$ and $\overline S_{\dot
\alpha }^I$ can also be easily written, 
roughly by simply replacing $\eta ^\alpha _I$ by $\eta
^\alpha _I + x^{\alpha \dot \alpha}\overline \xi _{\dot \alpha I}$,
and similarly for $\overline \eta ^{I\dot \alpha}$ in \susyvar, but we
will not need these transformations here, as it is the action of 
$Q^I_\alpha$ and $\overline Q_{I\dot \alpha}$ which generate
descendents.

The issue now is how the supersymmetry generators $Q^I_\alpha$ and
$\overline Q_{I\dot \alpha}$ act on the gauge invariant operators,
which are traces of products of the fields in \susyvar.  Classically
this given simply by the acting with the transformations in \susyvar\
on each of the fields in the operator.  For example, classical
expressions for some of the descendents of the short primary operator
$\O _2=[\Tr _G (\phi ^2)]_{(0,2,0)}$ are
\eqn\oiid{\matrix{&\hbox{operator}&SO(4) & SU(4)_R & U(1)_Y\cr
&\d O_2=\Tr (2\phi _{IJ} \psi_{K\alpha}+\phi _{KJ}\psi
_{I\alpha}-\phi _{KI}\psi _{J\alpha})&(\half,0)&(0,1,1)&-1\cr
&\d ^2 O_2=\Tr (\psi _{\alpha I}\psi _{\beta J}\epsilon ^{\alpha
\beta} +g[\phi _{IK},\phi _{JL}]\phi
^{KL})&(0,0)&(0,0,2)&-2\cr 
&\d ^2 O_2=\Tr (\phi _{IJ}F_{\alpha \beta}+\psi _{I(\alpha}\psi
_{\beta)J})&(1,0)&(0,1,0)&-2.\cr}} (Irrelevant overall normalization 
factors are suppressed.)
Likewise, classical expressions for some of the descendents of the
long operator $O_K=\Tr _G(\phi _{IJ}\phi ^{IJ})$ are 
\eqn\okd{\matrix{&\hbox{operator}&SO(4) & SU(4)_R & U(1)_Y\cr
&\d O_K=\Tr (\phi ^{IJ}\psi _{\alpha J})
&(\half,0)&(1,0,0)&-1\cr
&\d ^2 O_K=\Tr (
g[\phi ^{IK},\phi ^{JL}]\phi
_{KL})&(0,0)&(2,0,0)&-2\cr 
&\d ^2 O_K=\Tr (\phi _{IJ}F_{\alpha \beta}+2 \psi _{I(\alpha}\psi
_{\beta)J})&(1,0)&(0,1,0)&-2.\cr}}

Generally, however, we must expect that the superconformal generators
have quantum corrections when acting on gauge invariant composite
operators.  For example, the dilatation generator $D$, which acts on
primary operators as $D=(-i)x^\mu \partial _\mu +\Delta$, clearly gets
quantum contributions when acting on composite operators because
$\Delta$, which gives the dimension of the operator, gets quantum
contributions.  Because $D$ appears in $\{Q_{\alpha} ^I, S^{\alpha}
_I\}$, the action of the supersymmetry generators on composite
operators clearly must also generally have additional quantum
contributions.

Because the short operators do not have quantum corrections to their
operator dimensions, it is also natural to expect that the classical
expressions for their operator descendents are, in fact, exact.  This
is compatible with \metricYc\ and Ward identities such as that
discussed in \KI\ applied to 2-point functions.  
On the other hand, we should generally expect that descendents of long
operators, such as \okd, do receive quantum corrections.  Indeed,
this is the resolution to the following ``puzzle'':

\subsec{A ``puzzle'' and comments about operator mixings}

Consider the two-point function
\eqn\sli{\ev{(\delta ^2 O_2)(x) (\delta ^2 O_K)(y)},}
involving the Lorentz spin $(0,0)$ operators in the $(0,0,2)$ and
$(2,0,0)$ of $SU(4)_R$, respectively; the first operator is given in
the second line in \oiid, while the classical expression for the
second operator is given in the second line in \okd.  If this
two-point function were non-zero, it would violate the $U(1)_Y$
selection rule; however, as indicated generally in \metsl, \sli\ must
vanish.  This follows {}from the conformal group because $\delta ^2
O_2$ is a conformal primary with exact dimension $\Delta =3$, while
$\delta ^2 O_K$ classically has dimension 3 but gets non-zero quantum
corrections, corresponding to the non-zero anomalous dimension of
$O_K$.  Since, for general non-zero $g_{YM}$, the two operators have
different dimensions, the conformal group requires that the two-point
function vanishes.  Indeed, the only operator which can have a
non-zero two-point function with $\delta ^2 O_2$ is the conjugate
short operator $\overline \delta ^2 O_2$, for which
\eqn\ssii{\ev{(\delta ^2 O_2)(x) (\overline \delta ^2
O_2)(y)}=- {2|G|\over (2\pi )^4|x-y|^6};} the result \ssii\ is known
to be exact since it is related by supersymmetry to a non-renormalized
current two-point function.

However, using the expressions in the second lines in \oiid\ and \okd,
we find a non-zero result for \sli\ at order $g_{YM}^2$ coming {}from:
\eqn\slii{\ev{(g_{YM}\phi ^3)_{IJ}(x)(g_{YM}\phi
^3)^{IJ}(y)}=C_2(G)|G|g_{YM}^2{1\over (2\pi )^6|x-y|^6}+O(g_{YM}^4),}
where $C_2(G)$ is the quadratic Casimir of gauge group $G$, normalized
to be $N$ for $SU(N)$, and the factor of $C_2(G)|G|$ comes {}from
$f^{abc}f_{abc}$.  

The resolution to this apparent puzzle is that there must be a quantum
correction to the second line in \okd\ which compensates for \slii,
preserving the vanishing of \sli.  To order $g_{YM}^2$, we must have
\eqn\okdii{\delta ^2 O_K\equiv L^{IJ}=
\Tr (g_{YM}[\phi ^{IK},\phi ^{JL}]\phi
_{KL})+\half g_{YM}^2{C_2(G)\over (2\pi )^2}\overline S^{IJ},} where
$\overline S^{IJ}=\overline \d ^2 O_2$ is the conjugate operator to
$\d ^2 O_2$ in \ssii.  Using \ssii, the $g_{YM}^2$ correction term in
\okdii\ cancels the contribution to
\sli\ from \slii.  At higher orders in $g_{YM}$ there can be
additional quantum corrections to \okdii.

Finally, we would like to comment on the issue of operator mixing.
Generally long primary operators need not be ``pure'' primaries, in
the sense that they need not be eigenvectors of the anomalous
dimensions matrix, which arises in the OPE with the dilatation
operator $D$.  It is the eigenvectors of this anomalous dimension
matrix, with differing eigenvalues, which are orthogonal in that
their two-point functions vanish.  Operators which are not
eigenvectors have non-zero two-point function mixings among
themselves; diagonalizing the two-point functions is a practical way
to obtain the pure primary eigenvectors.

One might thus be tempted to interpret the above puzzle differently:
rather than correcting the action of $\delta$ as in \okdii\ to make
\sli\ vanish, perhaps \sli\ is actually non-zero and simply expresses
that $\delta ^2 O_K$ is not an eigenvector of the anomalous dimension
matrix but, instead, mixes with other operators such as $\overline
S^{IJ}$?  This latter interpretation requires that 
\eqn\sliii{\ev{O _K(x)\Ot(y)}}
is also non-zero, as \sliii\ is related to \sli\ by supersymmetry.
The interpretation of the non-zero result for
\sliii\ would, similarly, be that 
$O_K= \Tr (\phi ^{IJ}\phi _{IJ})$ itself is not a pure primary
operator but mixes with other operators, including the operator
$\Otbar$, In other words, this interpretation would require that $O_K$
is actually superposition $O_K=\widetilde O_K+cg_{YM}^2
\Otbar +\dots$, where $\widetilde O_K$ and the other terms are
eigenvectors of the anomalous dimension matrix.  If this were the
case, the 2-point function \sliii\ would be given by $(const.)
g_{YM}^2/|x-y|^8$.  However, it is easily seen that this can not
happen in perturbation theory: because \sliii\ has classical scaling
dimension 6, perturbation theory can only lead to terms scaling as
$1/|x-y|^6$ up to additional perturbative corrections depending on
$\log(x-y)$.  Resumming the logs can lead to perturbative expressions
such as $f(g)/|x-y|^{6+O(g_{YM}^2)}$, but we do not expect to be able
to get the $1/|x-y|^8$ dependence above in perturbation theory.
Briefly put: we expect that, in perturbation theory, there can be
operator mixing only among operators with the same classical scaling
dimensions.

Since there is no other $SU(4)$ singlet
with classical scaling dimension $2$, we do not expect that the above
$O_K$ can have any operator mixing in perturbation theory and, in
particular, \sliii\ must vanish in perturbation theory.  Consequently,
we believe that \sli\ really must vanish and the correct
interpretation of the above puzzle is the one given above: that the
action of $\delta $ on long operators such as $O_K$ gets quantum
corrections.  

By this same argument, we expect that the perturbative quantum
corrections to the action of the supersymmetry generators on long
operators must also respect the classical scaling dimensions of
operators.  For example, $\delta ^2 O_K$ has classical scaling
dimension $3$ so there can be a quantum correction in perturbation
theory by an operator in the same $SU(4)_R$ representation which also
has classical scaling dimension $3$; this is compatible with \okdii.
Consider, on the other hand, $\delta O_K$, which is in the ${\bf 4}$
of $SU(4)_R$, with Lorentz spin $(\half , 0)$ and classical scaling
dimension $5/2$.  Because there is no other operator with the same
classical scaling dimension and Lorentz and $SU(4)_R$ representations,
we do not expect to find a quantum correction to $\delta O_K$ in
perturbation theory.  Thus, for example, we expect that
\eqn\sllast{\ev{(\delta O_K)(x)(\delta ^3 O_2)(y)}=0}
in perturbation theory, though we have not completed the task of
explicitly verifying this.  Again, the only way \sllast\ could be
non-zero is if $\delta O_K$ mixes with $\overline \delta ^3 O_2$, in
which case
\sllast\ would be proportional to $1/|x-y|^7$ -- but in perturbation
theory \sllast\ would go as $1/|x-y|^6$ up to $g_{YM}$ corrections in
$\log (x-y)$.

\newsec{Comments on instanton contributions to correlation functions}

An instanton of ${\cal N}=4$ super Yang-Mills, with arbitrary gauge
group $G$, has $8C_2(G)$ fermion zero modes, where $C_2(G)$ is the
Casimir of the adjoint representation, normalized to be $N$ for
$SU(N)$.  More generally, an instanton number $k$ configuration has
$8kC_2(G)$ fermion zero modes.  Of these fermion zero modes, $16$ are
special: they are 8 zero modes generated by acting on the instanton
configuration with the 8 supercharges $Q_{\alpha}^I$, and 8 generated
by the superconformal supercharges $\overline S^I_{\dot
\alpha}$.  The remaining 16 supercharges annihilate the instanton
(they generate the 16 fermion zero modes of the anti-instanton).  We
will denote these 16 special zero modes by $\lambda (x)$ and the
remaining $(8kC_2(G)-16)$ zero modes by $\chi (x)$.

The 16 $\lambda$ are exact zero modes, while the $\chi$ can generally
be lifted.  In particular, at the origin of the moduli space of vacua,
which is the vacuum of interest for conformal invariance, the $\chi$
zero modes can be lifted in multiples of 4 by a term in the instanton
action $S_{inst}=\dots + \chi ^4$; this is discussed in
\DHKMV\ and references therein.   An instanton contributes to a
correlation function only if the 16 zero modes $\lambda$ are soaked up
by the operators involved in a correlation function.  

The general procedure is to replace every operator in the correlation
function with its instanton background version, $O_i\rightarrow
O_i^{inst}$ by the prescription $\phi \rightarrow \phi ^{inst}$, $\psi
\rightarrow \psi ^{inst}$ and $F\rightarrow F^{inst}$, where $\phi
^{inst}$, $\psi ^{inst}$, and $F^{inst}$ are the adjoint scalars,
fermions, and self-dual field strength solutions in the instanton
backgrounds.  Expressions for these solutions for the general $SU(N)$
instanton background are quite complicated and can be found in \DHKMV.
(This uses the $ADHM$ construction, which is not known for exceptional
groups.) The instanton can then contribute to $\ev{\prod _i O_i}$ if
all 16 fermion zero modes $\lambda$ appear in $\prod _i O_i^{inst}$.
If the 16 $\lambda$ zero modes are indeed soaked up, it will always be
possible to soak up the remaining $\chi$ zero modes by bringing down
powers of $S_{inst}$.  We can thus just focus on the $\lambda$ zero
modes.

We will give a heuristic argument for a relation between the $U(1)_Y$
charge of an operator and how many $\lambda$ fermion zero modes it
contains.  Consider the supersymmetry relations \susyvar\ in an
instanton background, where we replace $F\rightarrow F^{inst}$.  The
solution $\psi^{inst}$ satisfying $\delta \psi ^{inst}= F^{inst}$ is
$\psi ^{inst}=\lambda F_{inst}$, and the solition $\phi ^{inst}$
satisfying $\delta ^2 \phi ^{inst}= F^{inst}$ is $\phi ^{inst}=\lambda
\lambda F^{inst}$.

Our basic observation is that the fermion zero mode $\lambda \sim
\delta ^{-1}$.  Thus, if an operator $O_{top}$ satisfies $\delta
O_{top}=0$, then $O_{top}$ contains no $\lambda$ fermion zero modes --
it can only depend on $F^{inst}$, and possibly also any of the
$(8kC_2(G)-16)$ $\chi$ zero modes.  If an operator $O_r$ has $\delta
^r O_r=O_{top}$, with $\delta O_{top}=0$, then $O_r$ has $r$ $\lambda$
fermion zero modes.

The operator
$\overline \delta$ annihilates the fields in the instanton background;
in an anti-instanton background the roles of $\delta$ and $\overline
\delta$ are reversed.  Assigning $\delta$ charge $-1$ under $U(1)_Y$
as in \KI\ (the sign is to agree with the supergravity convention for
the charges of the conjugate sources), all operators with $U(1)_Y$
charge $q>0$ vanish in an instanton background, as they are obtained
with $\overline \delta$s on a $U(1)_Y$ neutral primary.  Similarly,
only those operators with $U(1)_Y$ charge $q>0$ are non-vanishing in
an anti-instanton background.

Operators in short representations have $U(1)_Y$ charge $q$ with
$|q|\leq 4$.  In particular, the operator $\delta ^4 {\cal O}_p$ has
$U(1)_Y$ charge $-4$ and thus must be annihilated if acted on by
another power of $\delta$, $\delta ^5 {\cal O}_p=0$.  Thus $\delta ^ 4
{\cal O}_p$ can contain no $\lambda$ fermion zero modes, it can only
depend on $F^{inst}$ and the $\chi$.  More generally, a short operator
with $U(1)_Y$ charge $q$ has
\eqn\OSq{O_{S}^{(q)}\sim \lambda ^{4-|q|},}
in an instanton background, where the $\sim$ includes some polynomial
in $F^{inst}$ and $\chi$.

Similarly, operators in long representations have $U(1)_Y$ charge
$|q|\leq 8$.  An operator of the form $\delta ^8 O$ can thus have no
$\lambda$ zero modes in an instanton background, as it is annihilated
by $\delta$.  Thus, for a {\it generic} operator in a long
representation, 
\eqn\OLq{O_{L}^{(q)}\sim \lambda ^{8-|q|},}
in an instanton background, where again the $\sim$ includes 
some polynomial in $F^{inst}$ and $\chi$.  Of course there are some
long multiplets, obtained {}from $\Tr \phi ^p$ with $p<4$, which
truncate earlier on, i.e. the operator $O_{top}$ with $\delta
O_{top}=0$ has $U(1)_Y$ charge $|q|_{max}<8$.  An example is the
Konishi operator $O_K=\Tr (\phi ^{IJ}\phi _{IJ})$, for which $|q|
_{max}=4$.  As always, $O_{top}\sim \lambda ^{0}$ and thus the
generalization of \OLq\ to the other operators in the multiplet is 
$O_L^{(q)}\sim \lambda ^{|q|_{max}-|q|}$.

Consider a correlation function of $n$ operators in short
representations, $\ev{\prod _{i=1}^n {\cal O}_i ^{(q_i)}}$. Using
\OSq, the condition for instantons to contribute to the correlation
function is
\eqn\OSqs{\sum _{i=1}^n (4-|q_i|)=16,} 
in which case the $16$ $\lambda$ zero modes can be saturated.  It thus
follows that instantons (or anti-instantons) can contribute to a
given $n$-point function of operators in short representations
if and only if
\eqn\nshort{n=4+{|q_T|\over 4},}
where $q_T=\sum _{i=1}^n q_i$ is the total $U(1)_Y$ charge.  It thus
immediately follows that instantons can {\it never} 
contribute to $n<4$-point functions of short operators; this is
compatible with the conjectured non-renormalization of \refs{\LMRS,
\DFS} for $n\leq 3$ point functions.  We also see that instantons can
contribute to a $n=4$-point function only if $q_T=0$; this is
consistent with the conjectured selection rule \KI\ that $n\leq 4$
point functions with $q_T\neq 0$ exactly vanish.

\lref\green{M.B. Green, hep-th/9903124.}
We were not able to find a formula along the lines of \OSq\ in 
\DHKMV, but expect that it must be possible to prove, at least for
$SU(N)$, using the complete (and complicated) analysis presented
there.  Demonstrating
\OSq\ would provide further support for the matching between
instantons and $AdS_5\times S^5$ supergravity results found in \instr.
Indeed, the resulting relation \nshort\ nicely agrees with results
{}from $IIB$ string theory.  See, for example, sect. 3 of \green, where
the Grassman coordinates $\theta ^A$, $A=1, \dots 16$, correspond to
the 16 zero modes $\lambda$ of the (D)-instanton.  The supersymmetry
generators $Q_A=\partial/\partial \theta ^A$, which matches with
our basic observation that $\lambda \sim \delta ^{-1}$.  Relation
(3.7) of \green\ nicely corresponds to our \OSq\ and eqn. (3.14) of
\green\ corresponds to our \nshort.   
(This latter correspondence uses the fact that stringy interactions
involving $n$ fields only contribute to $n$-point functions; their
contribution to correlation functions with fewer operators vanishes
because the fields involved vanish when evaluated in the $AdS_5\times
S^5$ vacuum, as in \BGKRi.)  While much of the agreement found in
\DHKMV\ between multi-instanton collective coordinates and
$AdS_5\times S^5$ relied on large $N$ $SU(N)$, we expect that 
\OSq\ and \OLq\ apply for any gauge group.

Using \OSq\ and \OLq\ we also see that instantons can contribute to a
SSL three-point function, involving two short and one long operator,
only if $q_T=0$.  (More generally, for a $n$-point function involving
a generic long operator and $n-1$ short operators, instantons
contribute only if $n=3+\f{1}{4}|q_T|$.)  Therefore instantons are
nicely compatible with our conjectured selection rule \SSLYc.  We also
see that instantons can contribute to the two-point function of two
long operators only if $q_T=0$; this agrees with \metricYc.  (More
generally, instantons contribute to a $n$-point function involving two
generic long operators and $n-2$ short operators only if
$n=2+\f{1}{4}|q_T|$.)

\newsec{Perturbative checks of the selection rule for
$n\leq 4$ point functions.}

A non-trivial perturbative check of \SSSYc\ appears in \DFS, where the
leading order radiative corrections to a descendent correlation
function, which would violate \SSSYc\ if non-zero, was found to
vanish.  If one believes the conjectured \refs{\LMRS,
\DFS} non-renormalization of all 3-point functions of short operators,
the selection rule \SSSYc\ for 3 short operators would follow because
all correlation functions respect $U(1)_Y$ in the $g_{YM}\rightarrow
0$ limit \KI.  Because checks of \SSSYc\ for other $(SSS)$ descendent
3-point functions are similar to the example considered in \DFS, we
will not present any additional examples.  Instead, in this section,
we will present a non-trivial perturbative check of the selection rule
\SSSSYc\ for 4-point functions of short operators.  In the next
section, we present checks of the selection rule \SSLYc\ involving two
short and one long operator.

We consider the 4-point function of short operators:
\eqn\ivpfi{\ev{\O ^{(-2)}_2(x_1)\O ^{(-2)}_2(x_2)\O _2 (x_3) \O _2
(x_4)},} where $O_2(x)=[\Tr(\phi ^2)]_{(0,2,0)}$ is the primary
operator with $U(1)_Y$ charge zero and $O_2^{(-2)}\equiv S_{(IJ)}$ is
its second descendent, which appears in \oiid\ and is a Lorentz
scalar in the $(0,0,2)$ representation of $SU(4)_R$.  According 
\SSSSYc, \ivpfi\ must exactly vanish, for any gauge group $G$, as it
violates the $U(1)_Y$ selection rule.  Note that the $SU(4)_R$ group
theory does allow for a non-zero result for \ivpfi, so the vanishing
is non-trivial.

We carried out the perturbative calculation in $N=1$ component
language, where only $SU(3)$ subgroup of $SU(4)$ R-symmetry is
manifest.  Specifically, for the operators $O ^{(-2)}_2$ we took
flavor combinations $S^{11}$ and $S^{44}$ defined in Ref.~\DFS, while
$\O _2 (x_3)=(\bar{z}_1)^2$ and $\O _2 (x_4)=\bar{z} t z$. We denote
the $N=1$ scalar fields by $z$, while $t$ is a traceless $SU(3)$ flavor
generator.

\midinsert
\epsfxsize=5in \epsfbox{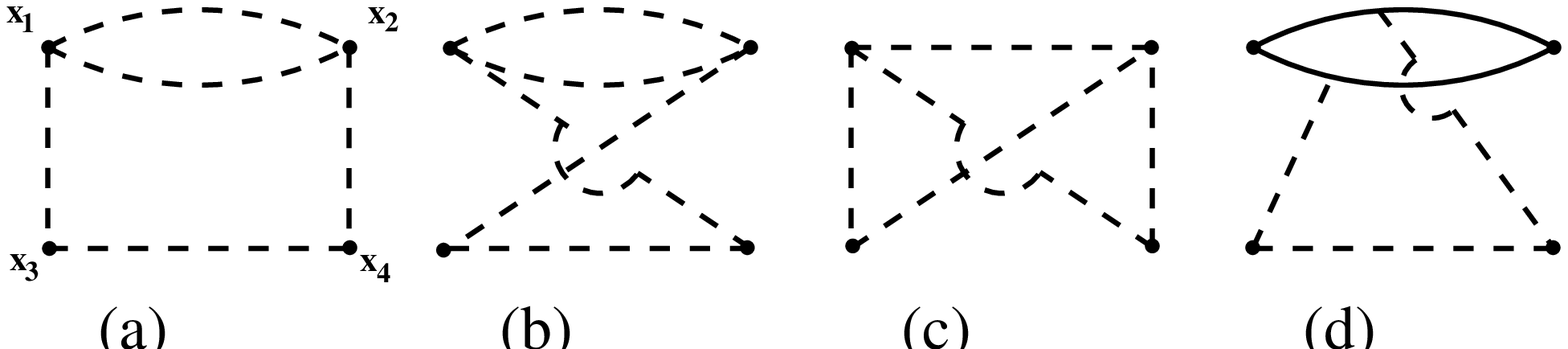} 

\noindent {\bf Figure 1.} Diagrams contributing to 4-point function
\ivpfi\ at $O(g^2)$. Dashed lines denote scalar propagators, 
solid ones fermion propagators.
\endinsert

There are 4 diagrams contributing to the 4-point function \ivpfi, they
are depicted in figure 1. The diagrams with exchanges of only scalar
fields are straightforward to evaluate and give
\eqn\bosonic{\eqalign{(a) &= g^2 C_2(G)|G| \, t_{11} \, 
G(x_1,x_2)^2 G(x_1,x_3) G(x_2,x_4) G(x_3,x_4) \cr (b)&= g^2 C_2(G)|G|
\, t_{11} \, G(x_1,x_2)^2 G(x_1,x_4) G(x_2,x_3) G(x_3,x_4)\cr (c)&=g^2
C_2(G)|G|(t_{22}+t_{33}) G(x_1,x_2) G(x_1,x_3) G(x_1,x_4) G(x_2,x_3)
G(x_2,x_4),}} where $G(x,y)=1/(4 \pi^2 |x-y|^2)$ is the free scalar
propagator and the factors of $C_2(G)|G|$ arise {}from factors of
$f^{abc}f_{abc}$.

Evaluating the diagram in fig.\ 1(d) involves an eight-dimensional
integral.  We found it easiest to perform such integrals in coordinate
space using the technique of conformal inversion \BJEF
\eqn\fermionic{\eqalign{(d)=(-) g^2 C_2(G)|G|\, t_{11} \, \big[&
G(x_1,x_2)^2 G(x_3,x_4) \left( G(x_1,x_3) 
G(x_2,x_4) + G(x_1,x_4) G(x_2,x_3)
\right)
\cr &- G(x_1,x_2) G(x_1,x_3) G(x_1,x_4) G(x_2,x_3) G(x_2,x_4) \big]. }}
The minus sign in front of the fermion contribution is the usual
fermion loop factor. Contributions {}from diagrams $(a)$ and $(b)$
cancel against the first two terms of the fermionic diagram. The
remaining part of diagram (d) and diagram (c) are proportional to
\eqn\trace{g^2 C_2(G)|G| \, (t_{11} + t_{22} + t_{33}),} 
which vanishes, for any gauge group $G$, since the $SU(3)_F$ generator
t is traceless.

\newsec{Perturbative checks of the $(SSL)$ selection rule.}

We now turn to some checks of the conjectured selection rule 
\SSLYc.  As discussed in sect. 4, there can be quantum
corrections to descendents of long operators.  To avoid this subtlety,
we first consider the situation where the long operator $O_L$ in
\SSLYc\ is primary and the short operators are descendents.

For our long primary operator, we take $O_L=\Tr _G(\phi
^{r+2s})_{(0,r,0)}$ where $s>0$ for this to be a long operator and the
subscript gives the Dynkin indices of the $SU(4)_R$ representation.
We can consider, for example, the 3-point function $\ev{O_L\ \delta \O
_p\ \delta O_q}$, for which $SU(4)_R$ allows a non-zero result
provided $r=p+q-1\ (mod\ 2)$ in the range $p+q-1\geq r\geq |p-q|+1$.
The leading contribution to this 3-point function, which would violate
$U(1)_Y$ conservation if non-zero, occurs at order $g_{YM}$ and is
associated with a single diagram, of the form of diagram (a) in
fig. 2.  Fortunately this diagram vanishes because it involves a color
contraction of the form: $f^{abc}d_{e_i(bc)}$, where $f^{abc}$ is
associated with the Yukawa interaction and $d_{e_i(bc)}$ is associated
with $O_L$, which appears in the diagram at the vertex without a
fermion line.  The sum vanishes due to the antisymmetry of $f$ and
symmetry of $d$ in the summed color indices $ab$.  We refer to such
vanishing as the $d\cdot f=0$ rule.

\midinsert
\epsfxsize=4in \epsfbox{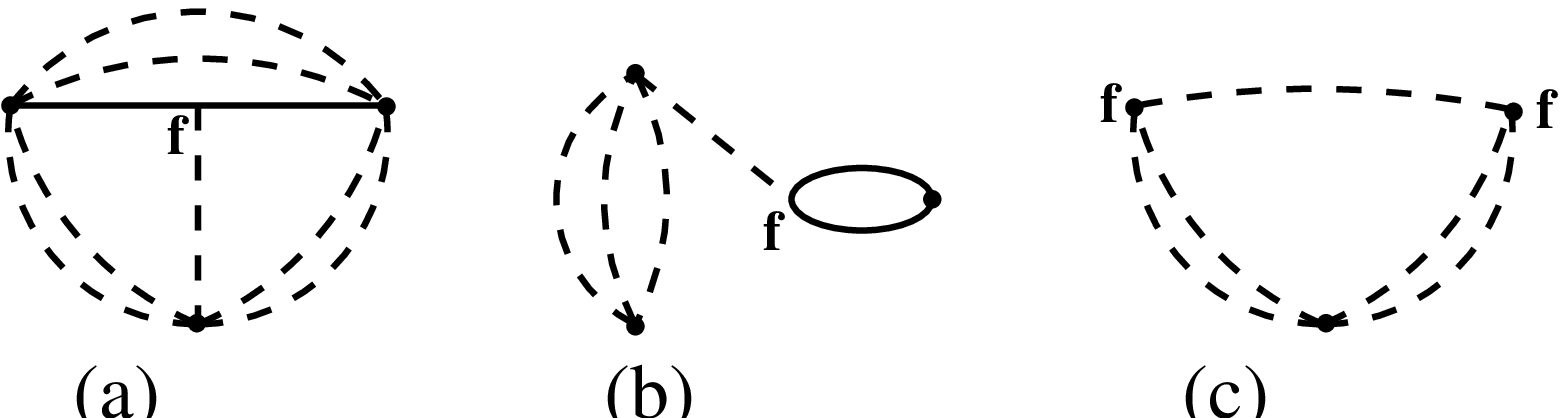} 

\noindent {\bf Figure 2.} Examples of contributions to the three-point
function \SSLYc . All these diagrams vanish due to contractions of the 
color indices.
\endinsert

We have thus verified that the leading radiative corrections to
$\ev{O_L\ \delta \O_p \ \delta \O _q}$ vanish, in agreement with
\SSLYc.  The task of
verifying that radiative corrections continue to vanish to higher
orders in $g_{YM}$ appears to be quite complicated and tedious, and we
have not carried it out.

We note that $SU(4)_R$ does not allow for a non-zero 3-point function
of the form $\ev{O_L \O _p \delta ^2 \O_q}$, with primary $O_L$.  So,
for our next examples, we consider $\ev{O_L\ \delta ^2 \O_p\ \delta ^2
O_q}$, where both $\delta ^2 \O _{p,q}$ are either the $(0,0)$ Lorentz
or $(1,0)$ Lorentz spin descendents, as in \oiid.  Consider first the
case where both are the Lorentz spin $(0,0)$ descendents.  Then
$SU(4)_R$ allows for a non-zero result if $p+q-2\geq r\geq |p-q|+2$,
with $r=p+q\ (mod\ 2)$.  The leading contribution to this 3-point
function, which would violate $U(1)_Y$ if non-zero, occurs at order
$g_{YM}^2$ and is associated with the diagrams (b) and (c) in fig. 2.
These indeed vanish by the $d\cdot f=0$ rule, where again the $d$
color factor is associated with the $O_L$ vertex.

The leading order contribution to $\ev{O_L\ \delta ^2 \O_p\ \delta ^2
O_q}$, where both $\delta ^2 \O _{p,q}$ are the $(1,0)$ Lorentz spin
descendents, is at order $g_{YM}^2$.  The relevant diagrams involve
gauge field propagators, which seem to complicate matters, and we did
not complete the task of evaluating the diagrams and verifying that,
as required by \SSLYc, they indeed sum to zero.

Other cases for which the leading radiative contributions can be
easily verified to vanish are $\ev{\delta O_L\ \delta \O _p\ \O _q}$
and $\ev{\delta O_L\ \delta \O_p\ \delta ^2 \O _q}$.  The relevant
diagrams are again of the types shown in fig. 2 and, in both cases,
they vanish by the $d\cdot f=0$ rule.  Evaluating the leading
radiative correction for $\ev{\delta O_L
\ \delta ^3 \O _p\ \O _q}$ is more difficult, as the diagrams  
involve gauge field propagators and, again, have not completed this
task.

Finally, we consider an example involving a second descendent of a
long operator, where the quantum corrections found in \okdii\ will
prove crucial.  Consider $\ev{\delta ^2 O_K\ \delta O_2 \delta O_2}$,
which has the $SU(4)_R$ flavor structure
$L^{IJ}S_{AB,I}S_{CD,J}\epsilon ^{ABCD}$.  We consider
$L^{44}S_{12,4}S_{34,4}$ in $N=1$ component fields where, using
\okdii, $L^{44}=\Tr \big(g_{YM}
\overline z^1[\overline z^2,\overline z ^3]+{1\over 8 \pi^2} g_{YM}^2
C_2(G)(\overline \lambda \overline \lambda +g_{YM}
\overline z^1[\overline z^2,\overline z ^3])+\dots \big)$,
$S_{12,4}= 
2 \overline z ^3 \lambda - z_2 \psi_1 + z_1 \psi_2$, and $
S_{34,4}= 3 z_3 \lambda$.  

The leading contribution to the correlation function is at order
$g_{YM}^2$ and includes a term $g_{YM}^2C_2(G)|G|
G_F(x-y)G_F(x-z)G_B(y-z)$ at Born level, coming {}from the order
$g_{YM}^2$ correction in \okdii.  The other order $g_{YM}^2$ terms
come {}from the order $g_{YM}$ term in $L^{IJ}$ along with one
interaction vertex.  There are two identical contributions coming
{}from the $-z_2\psi _1$ and the $z_1\psi _2$ terms in $S_{12,4}$.
The sum of these two contributions precisely cancel the above
additional term associated with the correction in
\okdii.  This is a non-trivial check of our conjecture, as the
coefficient of the correction term in \okdii, which was precisely
right to cancel the radiative corrections found here, was
independently determined in sect. 4.

\newsec{Comments on $SL(2,Z)$ S-duality and the OPE}

It was conjectured in \KI\ that an arbitrary $n$-point function 
of short operators transforms under $SL(2,Z)$ modular transformations
as 
\eqn\genslz{\ev{\prod _i \O ^{(q_i)}_i(x_i)}_{{a \tau +b\over c\tau
+d}}= \left({c\overline \tau +d\over c\tau +d}\right) ^{q_T/4}
\ev{\prod _i \O ^{(q_i)}_i(x_i)}_{\tau},}
where $\tau \equiv {\theta _{YM}\over 2\pi}+4\pi i g_{YM}^{-2}$ and 
$q_T=\sum _i q_i$ the net $U(1)_Y$ charge of the correlation
function.  This conjecture was motivated by AdS duality, but could
apply generally for arbitrary gauge groups\foot{In the case of 
$Sp(n)$ and $SO(2n+1)$, which are exchanged
by $\tau \rightarrow -1/\tau$, the correlation functions on the two
sides of \genslz\ would be for these two dual groups.}.  It is natural
to expect that \genslz\  applies for any operator correlation function,
including correlation functions involving long operators.

The conjecture \genslz, applied for arbitrary long or short operators,
is compatible with an OPE expansion of correlation functions.  Since
$U(1)_Y$ charge is additive, it is consistent to associate the modular
transformation properties in
\genslz\ entirely with the modular transformation
properties of the OPE coefficients and metric.  By our $U(1)_Y$
selection rule, the metrics $\eta _{ij}$ \metricYc\ and
$C^{SSS}_{ijk}$ and $C^{SSL}_{ijk}$ OPE coefficients are expected to
be modular invariant under $SL(2,Z)$ transformations of $\tau$.  As
discussed above, according to our conjectures $\eta ^{SS}_{ij}$ and
$C^{SSS}_{ijk}$ are actually constants independent of $\tau$, while
$\eta ^{LL}_{ij}$ and $C^{SSL}_{ijk}$ are non-trivial functions of
$\tau$, which should nevertheless be modular invariant.  In order to
satisfy \genslz\ the anomalous dimensions $\Delta _i$ of all operators
must be modular invariant; for short operators they are constant,
while for long operators they should be non-trivial, modular
invariant, functions of $\tau$.

The $U(1)_Y$ charge violation of general correlation functions is
associated entirely with the charge violation $(q_T)_{ijk}$ of the OPE
vertices $C^{SLL}_{ijk}$ and $C^{LLL}_{ijk}$.  Correspondingly, the
non-trivial modular transformation properties \genslz\ of a general
correlation function is associated entirely with the modular
transformation properties of $C^{SLL}_{ijk}(\tau)$ and
$C^{LLL}_{ijk}(\tau)$,
\eqn\Cslz{C^{XLL}_{ijk}({a\tau +b \over c\tau +d})=\left({c\overline
\tau + d\over c\tau + d}\right)^{(q_T)_{ijk}/4}C^{XLL}_{ijk}(\tau ).}
The general correlation function \genslz\ involves products of the
factors in \Cslz.

\centerline{{\bf Acknowledgments}}

We would like to thank D.Z. Freedman and A.V. Manohar for discussions.
This work was supported by UCSD grant DOE-FG03-97ER40546.  This work
was initiated while KI was on leave {}from UCSD and visiting the IAS,
where he was fully supported by an IAS grant {}from the W.M. Keck
Foundation; KI would like to thank the IAS for this support and 
for much hospitality.

\listrefs
\end